\documentstyle[12pt]{article}
\def\e{\mbox{e}}
\def\be{\begin{equation}}
\def\ee{\end{equation}}
\begin{document}

\begin{titlepage}
\begin{flushright}
INR-916e/96\\
gr-qc/9604038\\
\today\\
\end{flushright}

\begin{centering}
\vfill
{\bf A NEGATIVE MODE ABOUT EUCLIDEAN WORMHOLE}

\vspace{1cm}
V. A. Rubakov$^{\rm a,}$\footnote{rubakov@ms2.inr.ac.ru} and
O. Yu. Shvedov$^{\rm a,b,}$\footnote{olshv@ms2.inr.ac.ru} \\
\vspace{0.3cm}
{\em $^{\rm a}$ Institute for Nuclear Research of the Russian Academy of
Sciences,\\ 
60-th October Anniversary Prospect, 7a, Moscow, 117312, Russia\\}
\vspace{0.3cm}
{\em $^{\rm b}$ Physics Department, Moscow State University,\\
          Vorobyovy Gory, Moscow, Russia\\}

\vspace{0.7cm}

{\bf Abstract}

\end{centering}
\vspace{0.7cm}

Wormholes -- solutions to the euclidean Einstein equations with non-trivial
topology -- are usually assumed to make real contributions to amplitudes in 
quantum gravity. However, we find a negative mode among fluctuations about
the Giddings-Strominger wormhole solution. Hence, the wormhole contribution 
to the euclidean functional integral is argued to be purely imaginary
rather than real, which suggests the interpretation of the wormhole as
describing the instability of a large universe against the emission of
baby universes.

\vfill \vfill
\noindent

\end{titlepage}
\newpage

Euclidean wormhole configurations (fig. 1) 
may make non-trivial contributions
into the functional integral in quantum gravity. Originally it was
suggested
\cite{Hawking1,LRT1,GS1} that they may lead to loss of quantum coherence
for a macroscopic observer. It was then argued \cite{Coleman1,GS2} (see
also refs. \cite{Banks1,Banks2}) that the effects of wormholes on long
distance physics can be absorbed into the redefinition of $c$-number
coupling constants of the low energy theory, so that quantum coherence
is not lost. All this discussion depends crucially  on whether the
wormhole contributions into the functional integral are real or
imaginary. In particular, the arguments of refs. \cite{Coleman1,GS2}
rely upon the assumption that these contributions are purely real. On
the other hand, imaginary wormhole contributions would imply an
instability of the large universe with respect to the emission of baby
universes, in accord with the picture of refs. \cite{Hawking1,LRT1,GS1}.
The latter case is realized in a model which implements the ideas of
refs. \cite{Hawking2,Lyons}:  parent and baby universes are modelled in
$(1+1)$ dimensions by macroscopic and microscopic strings
\cite{VR,Nirov};
the wormhole contributions (string loops)
into the forward amplitudes 
are complex in this model, and it has been argued
\cite{VR} that the emission of baby universes leads
to the loss of quantum coherence in the parent $(1+1)$-dimensional
universe.

Clearly, it is hardly possible to decide whether generic wormhole
configurations in $(3+1)$ dimensions
make real or complex contributions into the functional integral. One may
try, however, to approach this problem semiclassically in models which
admit Euclidean wormhole solutions. By analogy to the analysis of
instantons/bounces in quantum mechanics and field theory
\cite{Langer,ColemanBounce,CallanColeman}, the
wormhole contibution will be imaginary if there exists a negative mode
among fluctuations about the classical solution\footnote{The analogy
to field theory suggests that the decay interpretation would require the
existence of exactly one negative mode \cite{ColemanNP}. It is
not straightforward to
see, though, how this requirement would emerge 
in quantum gravity context.}. The purpose
of this paper is to show that there exists a negative mode about the
simplest solution, the Giddings--Strominger \cite{GS1} wormhole.

The model with the Giddings--Strominger wormhole contains space-time
metrics and anti-symmetric tensor $H_{\mu\nu\lambda}$ as field
variables. As far as the wormhole dynamics is concerned, the equivalent
formulations \cite{Lee,AbbottWise,ColemanLee} are provided by using,
instead of $H_{\mu\nu\lambda}$, either conserved current
$J^{\mu}(x)$ or axion field $a(x)$,
\be
      J_{\mu} = \partial_{\mu} a
\label{1*}
\ee
The formulation convenient for our purposes is one in terms of $J^{\mu}$.
The Euclidean action is the sum of the action of pure gravity (the 
Hilbert--Einstein action
with boundary terms\footnote{The
boundary terms will play minor role in what follows.}) and
\be
     S_{matter}=f^2\int~d^4x {\sqrt{g}} g_{\mu\nu}
                 J^{\mu}J^{\nu}
\label{2*}
\ee
with the instruction that the functional integration is performed over
conserved current densities (see ref. \cite{ColemanLee} for details),
\be
     \partial_{\mu}(\sqrt{g} J^{\mu}) = 0
\label{2**}
\ee
$f$ in eq.(\ref{2*}) is the coupling constant.

The wormhole solution \cite{GS1} is $O(4)$-symmetric. In this paper we
consider only $O(4)$-symmetric fluctuations about this solution. The
general $O(4)$-symmetric Euclidean metrics is
\[
      ds^2 = N^2(\tilde{\rho})d\tilde{\rho}^2 + 
          \tilde{R}^2(\tilde{\rho})d\Omega^2
\]
where $d\Omega^2$ is the metrics on a unit 3-sphere. The
$O(4)$-symmetric current density has one non-zero component,
$J^0(\tilde{\rho})$, and its conservation, 
eq.(\ref{2**}), means that $J^0 \sqrt{g}$ is a
constant independent of $\tilde{\rho}$. 
This constant is related to the global
charge $Q$ flowing through the wormhole, $Q=2\pi^2 \tilde{R}^3 N J^0$. 
The action for
the $O(4)$-symmetric fields is then
\be
     S=\frac{3\pi}{4}\left(M_{Pl}L\right)^2
        \int~d\rho~ \left(-\frac{R}{N} R'^{2} - NR
                +\frac{N}{R^3}\right)
\label{3*}
\ee
where prime denotes derivative with respect to $\rho$, 
\[
     L^4 = \frac{2f^2Q^2}{3\pi^3 M_{Pl}^{2}}
\]
is a fixed  parameter for given wormhole type (i.e., for given $Q$), 
and we scaled this parameter out by introducing variables
\[
       R=\frac{\tilde{R}}{L},~~~~~~
       \rho=\frac{\tilde{\rho}}{L}
\]
One may hope that the semiclassical analysis is relevant at 
$L\gg M_{Pl}^{-1}$.

The metrics of the wormhole solution \cite{GS1} has $N_c=1$, while 
$R_c(\rho)$
obeys the only non-trivial equation following from the action
(\ref{3*}),
\be
     R'^{2}_c = 1 - \frac{1}{R^4_c}
\ee
The metrics becomes flat at large $|\rho|$ 
(i.e., $R_c(\rho) \to |\rho|$ as $|\rho| \to \infty$). The origin of the
coordinate $\rho$ can be chosen in such a way that $\rho=0$ corresponds
to the minimum size of the wormhole (turning point, $R'_c=0$); this size
is equal to $\tilde{R}_c (\rho=0) = L$.

Let us now consider $O(4)$-symmetric fluctuations about the wormhole
solution, i.e., set 
    $ R(\rho) = R_{c}(\rho) + r(\rho)$, $N(\rho) = 1 + n(\rho)$
and evaluate the quadratic in $(r,n)$ part of the action (\ref{3*}).
This quadratic action is invariant under the gauge transformations
($O(4)$-symmetric general coordinate transformations)
$n \to (n+ \xi')$, $r \to (r + R_{c}' \xi)$, where $\xi(\rho)$ is the gauge
function. Non-gauge modes can be chosen to satisfy $n(\rho) = 0$, while
$r(\rho)$ is not subject to any constraint (for the discussion of
admissible gauges in a similar context see ref. \cite{Sasaki}).
 This
gauge will be chosen in what follows.
Before writing down the quadratic action for $r(\rho)$, 
we notice that this action is unbounded from below, because of
the negative sign of the derivative term. This is the usual
problem with fluctuations of the scale factor, and it is cured by
performing the rotation \cite{GHP} $r \to i r$ (for further discussion
of the rotation of the conformal factor see refs.
\cite{Mazur,Blau,Mottola}). After making this rotation one obtains
\be
     S^{(2)}[r] = \frac{3\pi}{4} (M_{Pl}L)^2
                 \int~ d\rho~ \left(  R_{c} r'^{2} -
                \frac{8}{R_{c}^{5}} r^2\right) + 
                 \mbox{boundary~~terms}
\label{6*}
\ee
Let us see that this action has one negative mode.

The argument is the usual one \cite{CallanColeman}. There exists a zero
mode, $r^{(0)} = R_{c}'$, which is the translational zero mode remaining
after gauge fixing (it corresponds to the gauge parameter $\xi$
independent of $\rho$). This function has a node, so there exists a
negative energy ground state of the corresponding Schr\"{o}dinger
operator, which is the negative mode of the action (\ref{6*}). 
The fact that the determinant of the $O(4)$-symmetric fluctuation
about the wormhole solution is imaginary, can be understood also
on more general grounds within Maslov's theory \cite{Maslov1,Maslov2}.
However, one might
worry that the zero mode does not vanish at large $|\rho|$ because
$R_c(\rho) \to |\rho|$ (though this peculiarity can be dealt with by an
appropriate choice of the integration measure for the corresponding
quantum mechanical problem). One might also wonder whether the boundary
terms in eq. (\ref{6*}) play any role. So, we present here the explicit
check that there exists exactly one $O(4)$-symmetric
negative mode, and find its form.

Let us consider the eigenvalue equation for fluctuations, which
diagonalize the action (\ref{6*}),
\be
      R_c\left[ -\left(R_c r'\right)' - \frac{8}{R_c^5} r \right] =
           \omega^2 r
\label{7*}
\ee
The overall $\rho$-dependent factor multiplying the left hand side is
arbitrary, and we have chosen it in such a way that eq. (\ref{7*}) can
be solved explicitly (this is the same trick that works nicely in the
calculation of determinants about the Yang--Mills instanton \cite{tHooft}).
We are interested in negative $\omega^2$. Upon introducing a new variable
$y = R_c^{-4}(\rho)$ instead of $\rho$, and writing
$r=R_{c}^{-|\omega|}\psi(y)$, one rewrites eq. (\ref{7*}) as follows,
\be
  y(1-y) \frac{d^2 \psi}{d y^2} +
  \left[\left(1 + \frac{|\omega|}{2}\right) 
   - \left(\frac{3 + |\omega|}{2}\right) y \right] \frac{d\psi}{dy}
    + \left(\frac{1}{2} - \frac{|\omega|}{8} 
        +\frac{\omega^2}{16}\right) \psi
              =0
\label{8+}
\ee
The variables $y$ and $\rho$ are not in one to one correspondence.
This can be dealt with by requiring that $r(\rho)$ is either symmetric or
anti-symmetric in $\rho$. As $(R_c(\rho) -1)$ is symmetric, the latter
requirement means that $\psi(y)$ is either a series in $(1-y)$
(symmetric eigenfunctions of eq. (\ref{7*})) or a series in odd powers
of $\sqrt{1-y}$ (anti-symmetric eigenfunctions) at small $(1-y)$. 
We also impose the
condition that the eigenfunctions $r(\rho)$ are square integrable with
the weight $d\rho/R_{c}(\rho)$, which is appropriate for the choice of
the pre-factor made in eq. (\ref{7*}). In terms of $\psi(y)$, this means
 square integrability with the weight
$y^{|\omega|/2}dy/(y\sqrt{1-y})$.
In fact, the precise conditions at $|\rho| \to \infty$ are not very
important; it is sufficient to require that $r$ is finite.

Equation (\ref{8+}) is the hypergeometric equation. It is
straightforward to see that there exists exactly one eigenfunction
obeying above conditions, which is 
\[
\psi=\mbox{const~~~~~~ with~~~~~} \omega^2 =-4
\]
Other eigenfunctions of eq. (\ref{7*}) with negative $\omega^2$ grow
either as 
$\rho \to \infty$ or $\rho \to -\infty$. Thus, the only negative mode
has the form
\[
       r^{(-)}(\rho) = \frac{1}{ R_{c}^2(\rho)}
\]
Making use of this expression, one can check that the boundary terms in
eq. (\ref{6*}) (which are proportional to $R_{c} rr'$ or $R_{c}' r^2$)
vanish.

It is worth pointing out that the irrelevance of the boundary terms
in the gravitational action is the property of our gauge $N(\rho)=1$.
In other gauges the boundary terms may not vanish, and may even determine
 the sign of the quadratic action for fluctuations about the wormhole 
solution. The latter case is realized, for example, in conformal gauge,
$N(\rho)=R(\rho)$. One can check that, when boundary terms are 
taken into account, the determinant of $O(4)$-symmetric fluctuations
 is imaginary in any gauge.

We conclude this paper by adding a few remarks.

i) The existence of the negative mode implies that the wormhole 
contribution into the functional integral is imaginary, which 
corresponds to the instability of the parent universe against
the emission of baby universes. This fits nicely to the observation
\cite{LRT2} that the analytical continuation in $\rho$ describes a
 baby universe evolving, after its birth, in its intrinsic (real) 
time towards the singularity $R=0$. The analogy to $(1+1)$-dimensional
model of refs. \cite{VR,Nirov} is obvious.

ii) The wormhole contribution into the functional integral apparently
has wrong dependence on the spatial volume $V$ of the parent universe
and the normalization time $T$: the integration over the positions of
the two ends of the wormhole ($x$ and $y$ in fig. 1) results in the 
factor $(VT)^2$. We think this is an infrared effect inherent in
theories with Goldstone bosons. In the limit of small wormhole size,
the contribution of the wormhole of global charge $Q$ into the 
vacuum--vacuum amplitude can be summarized as follows 
(cf. \cite{ColemanLee}), 
\[
 \int ~dx~dy~A_{Q} <\e^{iQa(x)} \e^{-iQa(y)}>
\]
where $A_{Q}$ is a {\em purely imaginary} factor (exponentially
suppressed at large $Q$ by the wormhole action). This integral is indeed 
proportional to $(VT)^2$, which is the reflection of the existence of
(almost) zero energy intermediate states with charge $Q$. In massive
theories these states will be absent, and the wormhole contribution
 will be proportional to the usual factor $(VT)$.

iii) In our analysis of the $O(4)$-symmetric fluctuations about
 the wormhole solution, an important ingredient was the 
Gibbons--Hawking--Perry rotation, $r\to ir$. Although this prescription
 works well in other cases of tunneling in quantum gravity,
an independent check that the wormhole contribution is indeed 
imaginary, is
desirable. A promising formalism in this regard is the
Wheeler--De Witt wave function approach. Also, the analysis of
$O(4)$-asymmetric fluctuations about the wormhole is necessary.
We hope to clarify these points in future publications.

The authors are indebted to T. Banks, Kh. Nirov and P. Tinyakov for
helpful discussions. This work is supported in part by INTAS grant
94-2352 and Russian Foundation for Basic Research grant 96-02-17449a.

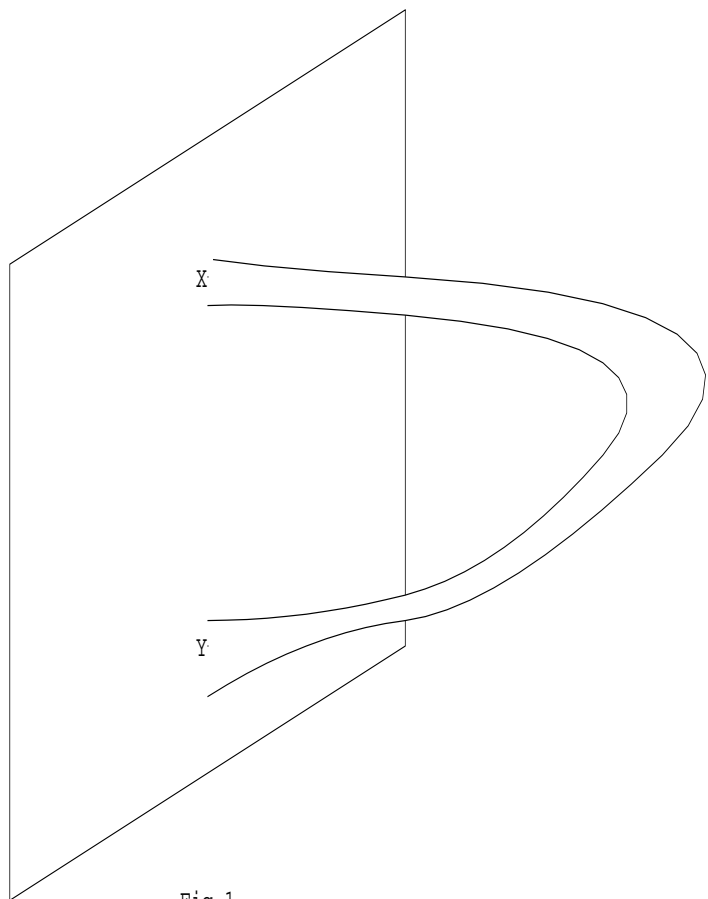
\end{document}